\documentclass[]{elsart}
\usepackage{graphicx}
\usepackage{amssymb,amsmath}
\usepackage[square,comma,numbers,sort&compress]{natbib}
\usepackage{amsfonts}
\usepackage{graphics}
\usepackage[thinspace,squaren]{mySIunits}

\def\cnlsq{{\textsc{cnlsq}}}
\def\dd{{\mathrm{d}}}
\def\drt{{\sf DRT}}
\def\hnef{{\sf HNEF}}

\def\full{\protect\mbox{------}}
\def\kesik{\protect\mbox{-\, -\, -}}
\def\kesiknoktali{\protect\mbox{-\, $\cdot$\, -}}

\def\ENcha{\protect\mbox{$\circ$}}
\def\ENchb{\protect\mbox{$\Box$}}
\def\ENchc{\protect\mbox{$\rhd$}}

\def\ENchdd{\protect\mbox{$\bigtriangleup\!\!\!\!\!\bigtriangledown$}}
\def\ENche{\protect\mbox{$\lhd$}}
\def\ENchf{\protect\mbox{$\star$}}
\def\ENchg{\protect\mbox{$\bigtriangledown$}}
\def\ENchh{\protect\mbox{$\bigtriangleup$}}

\journal{J Non-Cryst Solids}
\begin{document}
\begin{frontmatter}
  \title{Distribution of relaxation times in $\alpha$-phase polyvinylidene fluoride}
  \author{Enis Tuncer\corauthref{cor}},
  \corauth[cor]{Corresponding author.}
  \ead{enis.tuncer@physics.org}
  \author{Michael Wegener}, 
  \author{Reimund Gerhard-Multhaupt}
  \address{Applied Condensed-Matter Physics, Department of Physics, University of Potsdam, D-14469 Potsdam, Germany}

  \begin{abstract}
In this paper, a recently developed numerical method to analyze dielectric-spectroscopy data is applied to $\alpha$-phase polyvinylidene fluoride (PVDF). 
The numerical procedure is non-parametric and does not contain any of the extensively used empirical formulas mentioned in the literature. The method basically recovers the unknown distribution of relaxation times of the generalized dielectric function representation by simultaneous application of the Monte Carlo integration method and of the constrained least-squares optimization. 
The  relaxation map constructed after the numerical analysis is compared to $\alpha$-phase PVDF data presented in the literature and results of the parametric analysis with a well-known empirical formula. 
  \end{abstract}
\end{frontmatter}
Dielectric spectroscopy is a well established tool for materials characterization~\cite{KremerBook}. With this technique, the complex dielectric permittivity $\varepsilon(\omega)$ is obtained as a function of angular frequency $\omega$ ($\omega=2\pi\nu$; $\nu$ is the frequency of the applied voltage). Analysis of $\varepsilon(\omega)$ is usually performed with empirical formulas~\cite{Tuncer2000b,Macdonald2000a,MacDonald1987,Jonscher1983}. A neglected approach is the distribution of relaxation times (\drt), which is  based on functional analysis (it requires the solution of an integral equation)~\cite{FunctionalAnalysis}. The \drt\ expression is a Volterra equation~\cite{Volterra}, which is a special form of the Fredholm integral equations~\cite{Fredholm}. Such equations are usually considered to be {\em ill-conditioned} because of their non-unique solutions. However, the approach used here and recently presented elsewhere~\cite{Tuncer2004a,Tuncer2000b,TuncerSpectral,TuncerLic} leads to unique solutions by means of a constrained least-squares fit and the Monte Carlo integration method. ~\citet{macdonald:6241} stated that approaches with summations of delta functions would not suffer from the limitations of ill-posed inversions if complex nonlinear least-squares are employed, another approach to solve inverse problems. In this paper, we apply this numerical procedure that was developed  by one of us (ET) to a semi-crystalline polymer, $\alpha$-phase polyvinylidene fluoride ($\alpha$-phase PVDF)~\cite{TashiroFerro,JungnickelFerro,Grimau1999,Grimau1997}. An empirical formula and results from the literature are also included in order to evaluate the new \drt\ approach and to compare several analyzing techniques. Impedance measurements on $\alpha$-phase PVDF are performed at various tempatures $T$ between $173\ \kelvin$ and $450\ \kelvin$ within frequency windows of $0.1\ \hertz \le \nu \le 1\ \mega\hertz$.

\begin{figure}[t]
  \begin{center}
    \begin{minipage}[c]{0.50\linewidth}
      \includegraphics[width=\linewidth]{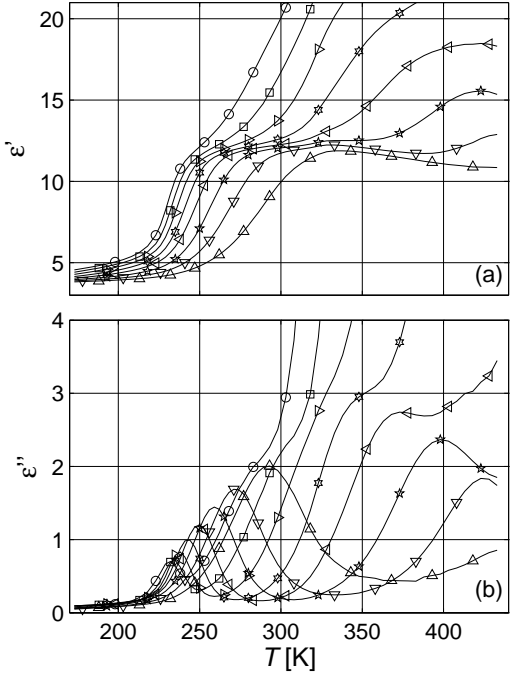}
    \end{minipage}\hfill
    \begin{minipage}[c]{0.48\linewidth}
      \caption{(a) Real and (b) imaginary parts of $\alpha$-phase PVDF dielectric function of temperature at various frequencies $\nu$: (\ENcha) $0.1\ \hertz$,  (\ENchb) $1\ \hertz$,  (\ENchc) $10\ \hertz$,  (\ENchdd) $100\ \hertz$,  (\ENche) $1\ \kilo\hertz$,  (\ENchf) $10\ \kilo\hertz$,  (\ENchg) $100\ \kilo\hertz$ and  (\ENchh) $1\ \mega\hertz$. The lines are drawn to guide the eyes. 
\label{fig:freq}}
    \end{minipage}
  \end{center}
\end{figure}
In Fig. \ref{fig:freq}, the relaxation spectra of $\alpha$-phase PVDF are shown at several frequencies as a function of temperature. Local movements of the structural units in $\alpha$-phase PVDF are observed at low temperatures, $T\lesssim250\ \kelvin$. This relaxation is usually called $\beta$ relaxation. In $\alpha$-phase PVDF, the magnitude of the $\beta$-relaxation is much smaller than those of two $\alpha$-relaxations. The $\alpha$-relaxations are related to cooperative movements of segments of the macromolecular chains. The subscripts '{\em a}' and '{\em c}' denote the fast asymmetric (amorphous) and the slow symmetric (crystalline) relaxation modes in the material~\cite{Grimau1999,Grimau1997}, respectively.  


The dielectric function of a material can be expressed in a general forma as follows,
\begin{eqnarray}
  \label{eq:1}
  \varepsilon(\omega)=\epsilon+\chi(\omega)+\sigma/(\imath\varepsilon_0\omega),
\end{eqnarray}
where $\epsilon$ and $\sigma$ are the frequency-independent relative dielectric permittivity and the ohmic conductivity of the material, $\imath=\sqrt{-1}$ and $\varepsilon_0=8.854\ \pico\farad\reciprocal\meter$. $\chi(\omega)$ is the susceptibility of the material which is usually expressed through the Havriliak-Negami empirical formula (\hnef)~\cite{HN} as
\begin{eqnarray}
  \label{eq:2}
  \chi(\omega)=\sum_j\Delta\varepsilon_j[1+(\imath\omega\tau_j)^{\alpha_j}]^{-\beta_j}.
\end{eqnarray}
Here, we consider that there are several relaxation processes in the material (summation over $j$). The parameters $\tau_j$ and $\Delta\varepsilon_j$ are the most probable relaxation time and the magnitude of the related polarization relaxation, respectively. The parameters $\alpha_j$ and $\beta_j$ determine the shape of the relaxation, $\alpha_j\le1$ and $\alpha_j\beta_j\le1$~\cite{TuncerPhysD}. A curve-fitting procedure based on the complex-nonlinear-least-squares (\cnlsq) method  
would yield the four unknowns for each relaxation $\{\tau,\,\Delta\varepsilon,\,\alpha,\,\beta\}$, as well as the  high-frequency dielectric permittivity $\epsilon$ and the ohmic losses proportional to $\sigma$~\cite{Macdonald2000a,MacDonald1987}. 

In the \drt\ representation, the susceptibility $\chi(\omega)$ of Eq.~(\ref{eq:1}) is expressed as
\begin{eqnarray}
  \label{eq:3}
    \chi(\omega)=\Delta\varepsilon\int_0^\infty {\sf g}(\tau)(1+\imath\omega\tau)^{-1} \dd\tau,
\end{eqnarray}
where ${\sf g}(\tau)$ is the distribution of relaxation times. Eq.~(\ref{eq:3}) is the Volterra equation~\cite{Volterra} and $(1+\imath\omega\tau)^{-1}$ its {\em kernel} $\mathcal{K}(\omega,\tau)$. For the numerical procedure, several relaxation times $\tau_k$ are selected from a log-linear scale on the frequency axis~\cite{Tuncer2000b,TuncerLic}. The number $k$ of the relaxations is smaller than the total number $n$ of experimental data points. In that case, the integral equation in Eq.~(\ref{eq:3}) becomes a sum that is linear in the ${\sf g}_k$, 
\begin{eqnarray}
  \label{eq:4}
    \chi(\omega)=\Delta\varepsilon\sum_k{\sf g}_k\ \mathcal{K}(\omega,\tau_k)
\end{eqnarray}
Since the ${\sf g}_k$ are expected to be positive, ${\sf g}_k\ge 0$, a constrained least-squares algorithm is employed~\cite{LawsonHanson,Adlers} with newly selected $\tau$ sets for $N$ times. $\alpha$-phase PVDF had previously been analyzed by ~\citet{Grimau1997} and by ~\citet{Grimau1999}. They used a Metropolis algorithm to obtain the relaxation characteristics of $\alpha$-phase PVDF.

\begin{figure}[t]
    \begin{center}
      \includegraphics[width=.9\linewidth]{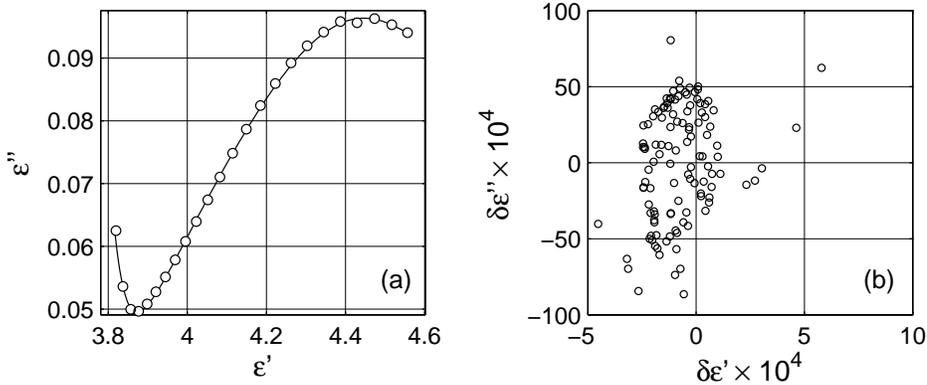}
      \caption{(a) Cole-Cole parametric plot of the dielectric function at $173\ \kelvin$. The symbols and the line represent the experimental data and the curve obtained from \cnlsq\ with the Havriliak-Negami empirical formula, respectively. (b) Relative error $\delta$ of the \cnlsq\ fitting.\label{fig:HN173}}
  \end{center}
\end{figure}

\begin{figure}[t]
  \centering
  \includegraphics[width=.9\linewidth]{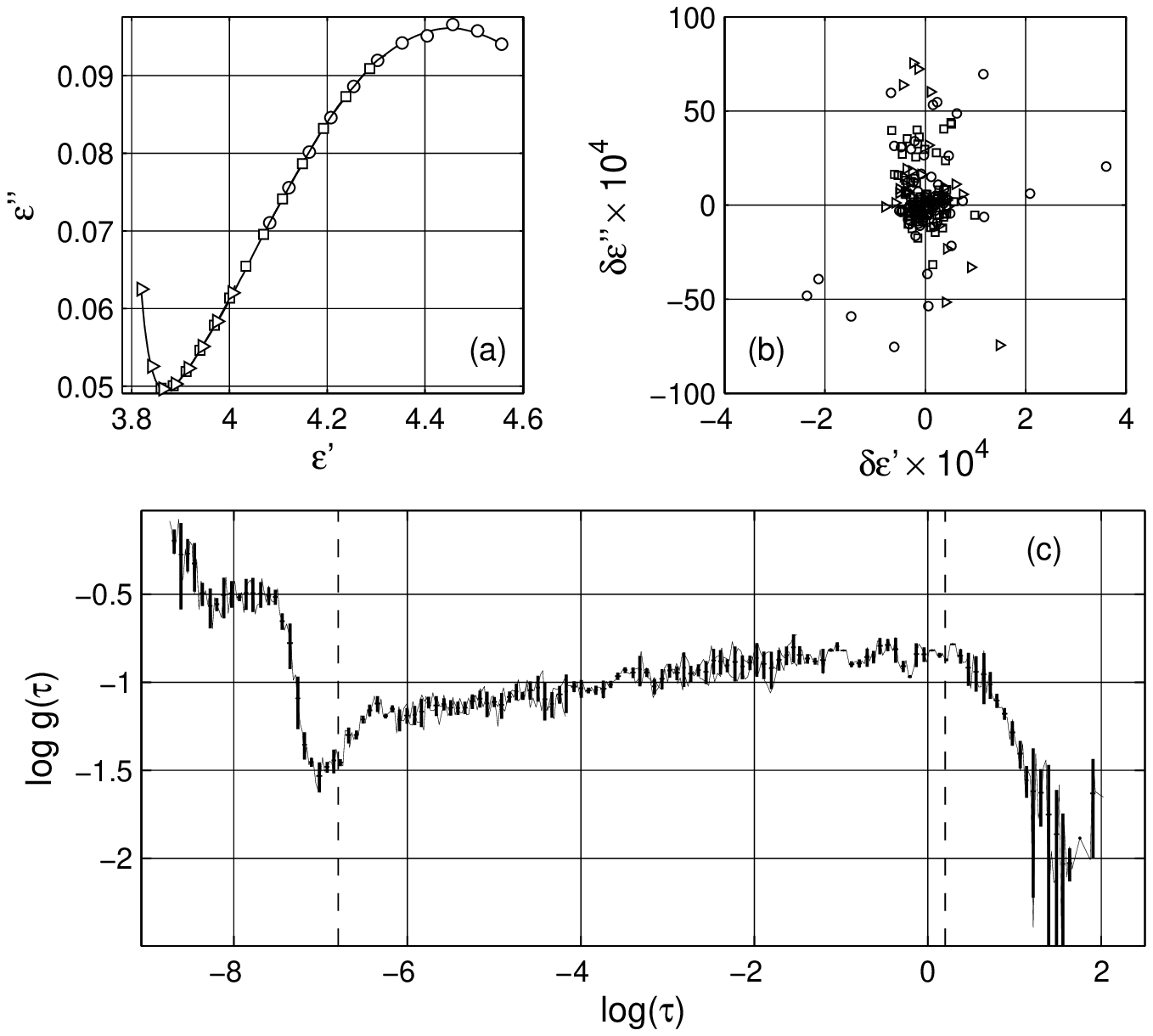}
\caption{(a) Cole-Cole parametric plot of the dielectric function at $173\ \kelvin$, the symbols and the line represent the experimental data and the curve obtained from the \drt\ analysis, respectively. The frequency window is divided into three regions on the frequency-axis, which are indicated with different symbols (`$\ENcha$';high, `$\ENchb$';intermediate and `$\ENchc$';low). (b) Relative error $\delta$ of the \drt\ analysis. (c) The \drt\ of $\alpha$-phase PVDF for the experimental window $0.1\ \hertz\le\nu\le1\ \mega\hertz$; the dashed vertical lines (\kesik) indicate the lower and upper limits of the relaxation times $\tau$ probed in the experiments.}
  \label{fig:DRT173}
\end{figure}

Below, we first present and compare the results obtained from \hnef\ and \drt\ analyses at $173\ \kelvin$. In Fig.~\ref{fig:HN173}, \hnef\ analysis is plotted. The symbols in Fig.~\ref{fig:HN173}a are the experimental values, and the line is the curve fitted with Eq.~(\ref{eq:2}) for $j=2$. The relative error $\delta$ of the real and imaginary parts of $\varepsilon(\omega)$ are presented in Fig.~\ref{fig:HN173}b. We illustrate the relative error in order to compare the \drt\ analysis with other methods. The average magnitude of the error in the imaginary part ${\delta\varepsilon''}$ is approximately ten times higher than that of the real part $\delta\varepsilon'$, $\delta\varepsilon''\simeq10\ \delta\varepsilon'$. In addition, a detailed analysis of the data shows that 90\% and 80\% of the data fall into the region $\delta\varepsilon'' <58\times10^{-4}$ and $\delta\varepsilon'' <48\times10^{-4}$, respectively.

In Fig.~\ref{fig:DRT173}, the \drt\ analysis based on Eq.~(\ref{eq:3}) is presented. Here, the data are divided into three regions (the results are presented in Fig.~\ref{fig:DRT173}a and \ref{fig:DRT173}b with open symbols). The number of relaxation times in each minimization step is chosen to be $2/3$ of the number of the experimental data points. The Monte Carlo loop $N$, which is the number of minimizations performed, is $2^{12}$. The average magnitude of the error in $\varepsilon''$ is again approximately ten times that of the real part, $\delta\varepsilon''\simeq\ 10\delta\varepsilon'$, as shown in Fig.~\ref{fig:DRT173}b. Compared to the \hnef, the \drt\ analysis yields lower relative errors.  90\% and 80\% of all the data points fall into the region $\delta\varepsilon'' < 40\times10^{-4}$ and $\delta\varepsilon'' < 26\times10^{-4}$, respectively. These values are nearly 50\% lower than the corresponding values of the \hnef\ analysis. In addition, the distribution of the errors is more symmetric than with the \hnef\ analysis. The extracted \drt\ spectrum is presented in Fig.~\ref{fig:DRT173}c, which is plotted with error bars after presenting the data in bins. The numerical procedure predicts smooth continuations of the relaxation times $\tau$ outside the experimental window. The experimental frequency window is indicated with dashed vertical lines (\kesik) in Fig.~\ref{fig:DRT173}c. 


\begin{table}[ht]
\caption{Fitting parameters for the $\alpha$ relaxations. $T_g$ is determined by the temperature at which the VFT fits yield $\log\tau=2$.\label{table}}
\begin{tabular*}{\linewidth}{@{\extracolsep{\fill}}lrrrrc}
\hline
  & $\log(\tau_0\ [\second])$ & $W\ [e\volt]$ & $T_0\ [\kelvin]$ & $T_g\ [\kelvin]$&Note\cr
\hline
\multicolumn{6}{c}{${\alpha_a}$ relaxation}\\
\hline
This work & $-13.6$ & 0.079 & $191.0$ &   223.4 &\drt\ \\
This work & $-12.8$ & 0.085 &  $191.3$ &  224.0 &\drt\ \\
This work & $-13.1$ & 0.076 &  $195.6$ & 225.4 &\hnef\ \\
~\citet{Grimau1999}& $-13.1$ &  0.088 & $187.5$ &  222.1& \hnef\ \\
\hline
\multicolumn{6}{c}{${\alpha_c}$ relaxation}\\
\hline
This work & $-16.6$ & 0.40 & \full  & \full  &\drt\ \\
This work & $-18.4$ & 0.45 & \full & \full &\hnef\ \\
~\citet{Grimau1999}& $-19.3$ &  0.49 & \full  & \full & \hnef\ \\
\hline
\end{tabular*}
\end{table} 

\begin{figure}[b]
  \begin{center}
    \begin{minipage}[c]{0.50\linewidth}
      \includegraphics[width=\linewidth]{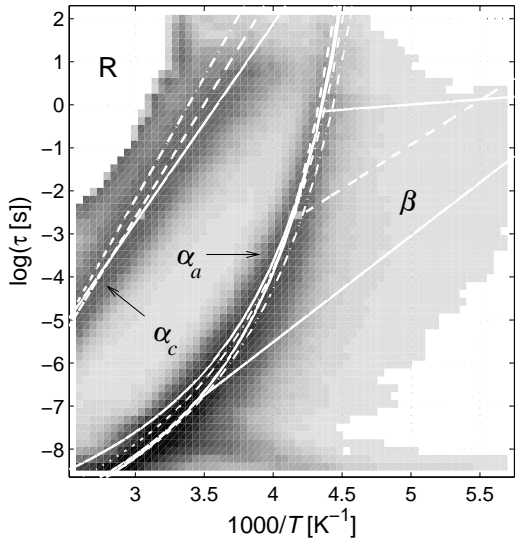}
    \end{minipage}\hfill
    \begin{minipage}[c]{0.48\linewidth}      
      \noindent\caption{Relaxation map for $\alpha$-phase PVDF, which clearly shows the $\alpha_a$ and $\alpha_c$ modes. Solid (\full) and dashed (\kesik) lines represent the fits to VFT or Arrhenius expressions for the  \drt\ and \hnef\ analyses, respectively. The results of ~\citet{Grimau1997} and ~\citet{Grimau1999} are plotted as chained lines (\kesiknoktali). The gray scale illustrates the probability density of the relaxations ${\sf g}(\tau)$, dark areas are the most probable regions for the particular relaxations. The omitted (white) region, labeled {\sf R}, would contain molecular information related to melting of the sample.\label{fig:relaxmap}}
    \end{minipage}
  \end{center}
\end{figure}

Application of the \drt\ analysis to data at different temperatures yields distributions similar to that presented in Fig.~\ref{fig:DRT173}c. L{\'e}vy statistics~\cite{Levy,BreimanLevy} will be later used to assess these distributions (see Refs.~\cite{Tuncer2004a,TuncerSpectral}). The mean relaxation times from the  L{\'e}vy statistics and from the \hnef\ analysis are employed to calculate the temperature dependence of the relaxations with the Vogel-Fulcher-Tammann (VFT) expression~\cite{Vogel,Fulcher,Tammann},
\begin{eqnarray}
  \label{eq:VFT}
  \tau=\tau_0 \exp\{W[k_b(T+T_0)]^{-1}\}
\end{eqnarray}
where $\tau_0$ is a constant, $W$ is a constant related to activation energy of the relaxation process, and $T_0$ is the Vogel temperature related to the glass transition $T_g$ of the material. For $T_0=0$, Eq.~(\ref{eq:VFT}) reproduces Arrhenius' expression, $\tau=\tau_0 \exp[W(k_bT)^{-1}]$ with $k_b=86.1321\ \micro e\!\volt\reciprocal\kelvin$. In Fig.~\ref{fig:relaxmap}, the relaxation map extracted from the \drt\ analyses at different temperatures is shown with gray scale. In order to compare both analyses, we plot the  \drt\ and \hnef\ results with solid (\full) and dashed (\kesik) lines, respectively.  Moreover, results from the literature~\cite{Grimau1999,Grimau1997} are also given [chained line (\kesiknoktali)]. The $\beta$ relaxation is omitted, since it has a very broad distribution and a low magnitude compared to the $\alpha$-relaxations as presented in Fig.~\ref{fig:DRT173}c. The fit parameters for the VFT and Arrhenius expressions and the values from the literature are presented in Table~\ref{table}. The $\alpha_a$ relaxation is known to be asymmetric, and our investigations by means of L{\'e}vy statistics confirm this observation---we are able to fit two symmetric relaxations that result in two different VFT expressions, see the solid lines (\full) in Fig.~\ref{fig:relaxmap}. The $\alpha_a$ distribution is symmetric at low temperatures, but becomes asymmetric around $250\ \kelvin$. When the fitting results are taken into consideration, there are slight differences between our studies and those of ~\citet{Grimau1999} and ~\citet{Grimau1997}, which are probably due to differences in the preparation and the thermal history of the materials. 

Here, we presented the results of dielectric-data analyses for $\alpha$-phase PVDF with two different techniques. For our particular material, both techniques result in equivalent temperature dependencies, which were also compared to data in the literature. Our numerical method for finding the \drt\ yields unique distributions for given dielectric data without any {\em a-priori} assumption.  We have thus demonstrated that the dielectric data of materials can be analyzed with a more general representation than the empirical formulas. The \drt\ method might be useful to investigate the origin of dielectric relaxations and the electrical properties of materials.


\end{document}